# N-DOPED SURFACES OF SUPERCONDUCTING NIOBIUM CAVITIES AS A DISORDERED COMPOSITE

W. Weingarten*, retiree from CERN, Geneva, Switzerland

## Abstract

The Q-factor of superconducting accelerating cavities can be substantially improved by a special heat treatment under $N_2$ atmosphere (N-doping). Recent experiments at Fermi National Laboratory investigated the dependence of Q on the RF frequency and showed, unexpectedly, both an increase and a decrease with the RF field amplitude. This paper shall explain this finding by extending a previously proposed model founded on the two fluid model of RF losses, percolation and the proximity effect in a disordered composite.

## INTRODUCTION

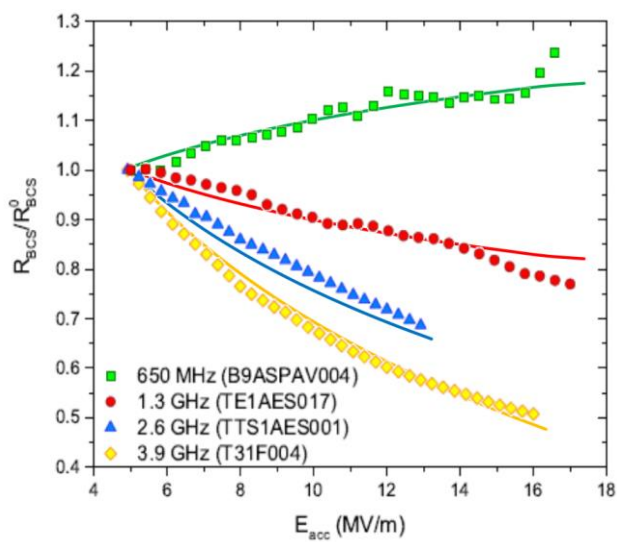

Figure 1: The surface resistance $R_s$ for N-doped cavities (made with the exact same doping recipe) as a function of the accelerating field $E_{acc}$ at 2 K, normalized to the low field surface resistance $R_s^0$ at 5 MV/m (adapted from ref. 9, where $R_s$ is called $R_{BCS}$). Superimposed are the results from eqs. (1) and (3), obtained with the parameters as in Table 1 (coloured lines), NB: $B/E_{acc}$=4mT/(MV/m).

The RF-field dependence of the Q-value (Q-slope) is an issue for the application of superconducting niobium cavities for high energy particle accelerators. Opposite phenomena were observed, extending from a decrease of the Q-value with RF field (Q-slope), in particular in copper cavities sputter-deposited with a thin niobium coating, to an increase of the Q-value with RF field (Q-rise). Theoretically plausible models were proposed for both of these observations [1,2,3,4,5], based on the dependence of the energy gap, i.e. the "BCS"-surface resistance $R_{BCS}$, on the RF field, or the reduction of the electronic mean free path by impurities or on the dynamic reduction or increase in number of quasiparticles by the action of the RF field. An assessment of these models is beyond the scope of this paper, but instead an alternative model is presented which postulates "weak" superconducting defects becoming normal with increasing magnetic field [6]. The beneficial effect of N-doping was discovered meanwhile [7,8].

In what follows the alternative model will be scrutinised under the recent experimental findings [9], which is the frequency dependence of the Q-value on the RF field (Figure 1). The approach chosen here is essentially valid for any "weak" superconductor in close proximity with a "strong" superconductor, such as niobium, but will be exemplified in what follows for nitrogen interstitially dissolved in niobium.

## RÉSUMÉ OF MODEL

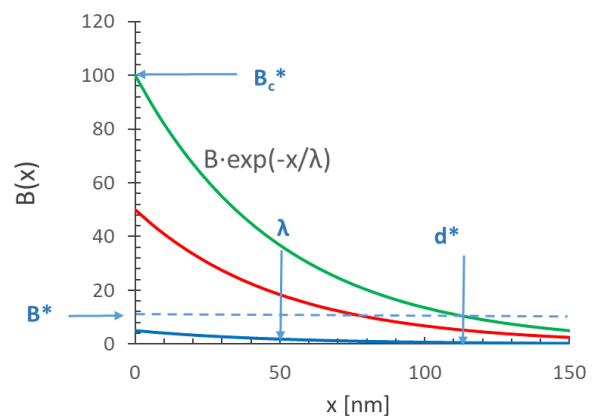

Figure 2: Decay of the magnetic field B(x) with the distance x from the surface; the "weak" superconductor extends to a depth $d^*$, which defines the saturation field $B_c^*$. Three different amplitudes of the RF magnetic field are drawn, the lowest one below the threshold field $B^*$, the larger one above $B^*$, and the largest one at the saturation field $B_c^*$.

The alternative model is built on the two-fluid description of the surface resistance $R_s$. The first component of the "weak" superconductor is the "nitrogen" component, consisting of nitrogen dissolved in niobium (volume fraction $x_1$). The second component of the weak superconductor is the "niobium" component, consisting of "dirty" niobium (volume fraction $x_2=1-x_1$). Depending on the concentration of nitrogen (~15 up to 25 at.%) the weak superconductor may have an intrinsic critical temperature with a lower bound of $T_{cN}$~1.2-2 K [10,11,12], but, by the proximity effect, have a larger one due to the coupling with the "dirty" niobium component. $T_{cN}$ is the critical temperature, where the nitrogen component itself becomes a superconductor. When both components of the weak superconductor are superconducting (at low RF field), the overall average electrical conductivity is named $s_{Nb}$.

---
* wolfgang.weingarten@cern.ch

When the RF magnetic field amplitude B is raised, the weak superconductor gradually turns normal conducting, consequent to the proximity effect, above a small threshold field $B^*$, up to a saturation field $B_c^*$. The threshold field $B^*$ is identical with the RF critical field of the weak superconductor. The saturation field $B_c^*$ describes the maximum RF field where all weak superconducting metal has turned normal conducting. It is linked with the depth $d^*$ of the weak superconductor $d^*=\lambda \cdot \ln(B_c^*/B^*)$. For increasing RF field from $B^*$ to $B_c^*$, the boundary normal conducting/superconducting penetrates deeper into the surface up to a maximum depth $d^*$, where the "weak" superconductor is vanishing ($\lambda$ is the penetration depth, Figure 2).

The low field surface resistance is $R_s^0$. For increasing B>$B^*$ some volume fraction f(B) no longer contributes to $R_s^0$, hence $R_s^0$ will be diminished by 1-f(B). This same fraction f(B) will instead acquire a different surface resistance $c \cdot R_s^0$. The constant c is the ratio of the average electrical conductivity of the weak superconductor $s_m$ including its two components and the overall average electrical conductivity $s_{Nb}$, $c=s_m/s_{Nb}$. Here the supposition is made of a volume fraction of the weak superconductor being constant within the depth of the RF field. This supposition is confirmed by surface analysis [6,13].

Hence the B dependence of $R_s$ can be described by

$$R_s = R_s^0 \cdot [1-f(B)+c \cdot f(B)] \quad . \quad (1)$$

Above $B^*$ the function $f(B)=\ln(B/B^*)/\ln(B_c^*/B^*)$ ($B^*<B< B_c^*$ and f(B)=0 for B<$B^*$ and f(B)=1 for B>$B_c^*$) describes the fraction of the weak superconductor that has become normal conducting (0<f(B)<1).

## EXTENSION OF MODEL

The model as described so far cannot explain the frequency dependence of $R_s$ with B. It has to be extended by a more detailed analysis based on a paper of R. Landauer [14,15], as already anticipated in a footnote of ref. 6.

Landauer, and Bruggeman already before him [16], considered a mixture of two metallic phases of individual conductivities $\sigma_1$ and $\sigma_2$, with $x_1$ and $x_2$ as their respective volume fractions (Effective Medium Approximation, EMA). Then the conductivity $\sigma_m$ of the infinite uniform medium is

$$4\sigma_m = (3x_1-1)\sigma_1 + (3x_2-1)\sigma_2 + \sqrt{[(3x_1-1)\sigma_1 + (3x_2-1)\sigma_2]^2 + 8\sigma_1\sigma_2} \quad . \quad (2)$$

The electrical conductivity of the "nitrogen" component of the weak superconductor, when normal conducting, is real, $\sigma_1=s_1$. The electrical conductivity of its "niobium" component is purely imaginary, $\sigma_2=(\mu_0\lambda^2\omega)^{-1} \cdot i=s_2 \cdot i$. This observation introduces the dependence on the frequency $\omega=2\pi f$ into eq. (2). Hence, the conductivity $\sigma_m$ of the weak superconductor is complex as well, and its real part $s_m$ describes its RF losses:

$$\mathrm{Re}(4\sigma_m) = 4s_m = (3x_1-1)s_1 + \left\{\left[(3x_1-1)^2 s_1^2 - (3x_2-1)^2 s_2^2\right]^2 + \left[2(3x_1-1)(3x_2-1)s_1 s_2 + 8s_1 s_2\right]^2\right\}^{1/4} \cdot$$

$$\cdot \frac{1}{\sqrt{2}}\sqrt{1+\frac{(3x_1-1)^2 s_1^2 - (3x_2-1)^2 s_2^2}{\sqrt{\left[2(3x_1-1)(3x_2-1)s_1 s_2 + 8s_1 s_2\right]^2 + \left[(3x_1-1)^2 s_1^2 - (3x_2-1)^2 s_2^2\right]^2}}} \quad . \quad (3)$$

## DATA ANALYSIS

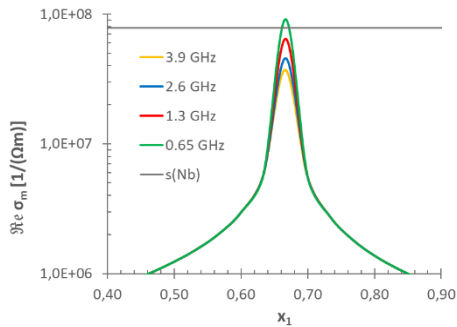

Figure 3: The real part of the electrical conductivity $\sigma_m$ of the weak superconductor vs the volume fraction $x_1$ of the "nitrogen" component. s(Nb) is the overall average electrical conductivity.

An impartial choice of parameters for fitting the data of Figure 1 is not obvious. Instead, it is observed that, within the proposed model, the relatively large dependence of the Q-value on the frequency is provided by a sharp peak of $\sigma_m$ located at a volume fraction around $x_1=0.667$ of the "nitrogen" component of the weak superconductor (Figure 3). This number is known to represent a percolation threshold in three dimensions for a binary metallic mixture in the EMA model.

As consequence, from ref. 12, the normal state conductivity of the "nitrogen" component amounts to $\sigma_1=s_1=5.5 \cdot 10^5$ $(\Omega m)^{-1}$. With $B^*$=20mT, as suggested by the data of Figure 1, the fitting procedure becomes straightforward: the residual resistivity ratio RRR at low field and the saturation field $B_c^*$ are the only free fitting parameters, all others are derived. These are the mean free path l, the penetration depth $\lambda$, the depth $d^*$ of the "weak" superconductor and the overall average electrical conductivity $s_{Nb}$ (cf. Table 1). Assuming a typical error of the data of ±3% ends up with $\chi^2$ compatible with the number of data points and free parameters. The error intervals of the fit parameters are defined at twice the minimum $\chi^2$.

One obtains finally $s_m$ from eq. (3) and $R_s/R_s^0$ from eq. (1).

It should be noted, however, that the conductivity $s_{Nb}$ is reduced by a factor $(T/T_c)^4=(2/9.2)^4\approx0.2\%$, by virtue of the temperature dependence of the two-fluid model. Keeping the constant c unchanged, as required by the data fitting, the average electrical conductivity of the weak superconductor $s_m$ is diminished by the same factor. This is a first indication of the weak superconductor being dispersed within the surface.

Table 1: Parameters used for fitting the data of Figure 1

| | | |
|---|---|---|
| $x_1$ | 0.667 | A priori parameters |
| B* [mT] | 20 | |
| l [nm] | 34±2 | Derived parameters |
| λ [nm] | 57±1 | |
| d* [nm] | 68±1 | |
| $s_{Nb}$ [1/Ωm] | (7.8±0.4)·10$^7$ | |
| RRR | 11.7±0.5 | Free fit parameters |
| $B_c$* [mT] | 66±5 | |
| London penetration depth $\lambda_0$ | 39 [nm] | |
| Intrinsic coherence length $\xi_0$ | 38 [nm] | |
| Mean free path l | 2.9·RRR [nm] | |
| Penetration depth λ | $\lambda_0\cdot(1+\xi_0/l)^{1/2}$ | |
| Depth of "weak" superconductor d* | $\lambda\cdot\ln(B_c^*/B^*)$ | |
| Overall average el. conductivity $s_{Nb}$ | RRR· $s_{Nb}$ (300K) | |
| El. conductivity of Nb@300 K $s_{Nb}$ (300K) | 6.7·10$^6$ [1/Ωm] | |

Consequent to the sharp value of $x_1$=0.667±0.002, the weak superconductor should have an atomic volume concentration of interstitially dissolved nitrogen of ~0.667·(15-25)%=(10-17)%, i.e. about every second atom in a line is a nitrogen atom, hence close to a composition such as NbN$_{1\pm x}$ with x«1. The hexagonal modifications of NbN and Nb$_2$N were found not to be superconducting down to 1.94K [17], in accordance to the assumptions of the proposed model. In addition, as a second hint of the weak superconductor being dispersed within the surface, the overall average atomic volume concentration of interstitially dissolved nitrogen as measured in N-doped cavities is more than one order of magnitude lower than 10%, about 0.5-1% [6,13].

## CONSISTENCY CHECK

In what follows the consistency of the results as obtained so far will be revised under the stipulations of the proximity effect as initially published by the Orsay group [18,19]. The properties of the "nitrogen" component of the weak superconductor are denoted by the suffix "N" for "normal", those of the "niobium" component are denoted with an "S" for "strong".

The starting point is the implicit formula eq. (1) in ref. 19 for the coherence length $K_N^{-1}$,

$$\ln(T_{cN}/T) = -\psi\left(\frac{1}{2}\right)+\psi\left(\frac{1}{2}-\frac{\hbar\cdot D_N\cdot K_N^2}{4\pi\cdot k_B\cdot T}\right).$$

Mind that this formula and the following ones taken from refs. 18 and 19 are written in the cgs-system of units.

$T_{cN}$ is the critical temperature of N itself, $\psi$ is the digamma function, $D_N$ is the diffusion coefficient, $\hbar$ is the Planck constant, $k_B$ is the Boltzmann constant, and T is the bath temperature.

The coherence length $K_N^{-1}$ depends implicitly on the diffusion constant $D_N=1/3\cdot v_{FN}\cdot l_N$, on $T_{cN}$=1.2K, and on T=2K. $v_N$ is the Fermi velocity of N, and $l_N$ is the electronic mean free path in N. The dependence of $l_N$ on RRR is $l_N$[nm]=(2.9±0.4)·RRR. This number is averaged from results of refs. 20,21 and 22. Then, by means of RRR=11.7 from the data fitting, $l_N$ is computed as 34[nm]. With $v_{FN}$=1.4·10$^6$[m/s] and RRR=11.8 from Table 1, $K_N^{-1}$=233[nm].

From eq. (3) of ref. 19, the penetration depth $\lambda_N$=57[nm] (Table 1) of the "weak" superconductor depends in addition on its electrical resistivity $\rho_N$ and on its energy gap $\Delta_N$.

The energy gap $\Delta_N$ was determined from the de Gennes-boundary condition, eq. (4) of ref. 19 $\Delta_N=\Delta_S\cdot NV_N/NV_S$. $\Delta_N$, $\Delta_S$ are the energy gaps and $NV_N$, $NV_S$ are the electron-phonon coupling constants of "N" and "S", respectively. The relation $T_c$=1.14·$\Theta_D\cdot e^{-1/NV}$, $\Theta_D$ being the Debye-temperature, allows the determination of the respective coupling constants, under the assumption of equal Debye temperature for "N" and "S". This approximation is considered as justified due to the logarithmic dependence of NV on $\Theta_D$.

All these numbers are known except $\rho_N$, which is adjusted such that $\lambda_N$=57[nm], as required from the data fitting. The numbers not yet mentioned so far are listed in Table 2.

From eq. (2) in ref. 19,

$$\xi_N = \left(\frac{\hbar\cdot D_N}{2\pi\cdot k_B\cdot T}\right)^{1/2} = \left(\frac{\hbar\cdot v_{FN}\cdot l_N}{6\pi\cdot k_B\cdot T}\right)^{1/2},$$

the coherence length $\xi_N$=97[nm] is derived, yielding the Ginzburg-Landau constant to $\kappa$=0.6.

Table 2: Parameters used for the consistency check

| $D_N$ [m$^2$/s] | $\Delta_N$ [K] | $\Delta_S$ [K] | $NV_N$ | $NV_S$ | $\Theta_D$ [K] | $\rho_N$ [Ωm] |
|---|---|---|---|---|---|---|
| 0.016 | 10.8 | 17 | 0.18 | 0.28 | 276 | 6.9·10$^{-8}$ |

Hence, the stipulated conditions of ref. 18, $\kappa$<1 and $d_N$»$\lambda_N$, allow the determination of the critical field of the "weak" superconductor from eq. (6) in ref. 18: $H_b\rightarrow\Phi_0\cdot K_N/(2\pi\lambda)$=24[mT], $\Phi_0$ being the flux quantum. $H_b$ should be consistent with the fitted value B*, which is the case. In addition, as clearly outlined in ref. 18, for increasing B>B*, the boundary wall normal-/superconducting penetrates into the surface like in a type I superconductor (NB $\kappa$<1), as already postulated in the above "résumé" section.

The fitted value of B* allows the determination of the "defect" size $d_N$ by virtue of eq. (10) in ref. 19: $d_N$=350[nm]:

$$H_{bN} = 3.8\frac{\Phi_0\cdot K_N}{2\pi\cdot\lambda}\cdot\exp(-K_N\cdot d_N).$$

## COMPARISON BETWEEN N-DOPED AND CHEMICALLY POLISHED CAVITY SURFACES

The authors of ref. 9 presented in comparison to Figure 1 also data for non N-doped superconducting cavities, which were chemically polished and baked at 120°C (Figure 4).

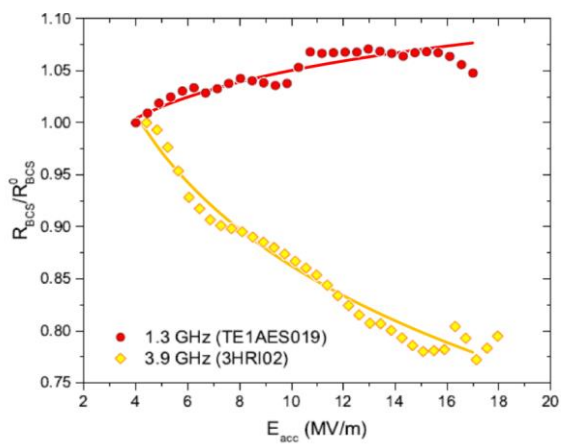

Figure 4: The surface resistance $R_s$ for buffered chemically polished (BCP) non-doped cavities as a function of the accelerating field $E_{acc}$ at 2 K, normalized to the low field surface resistance $R_s^0$ at 5 MV/m (adapted from ref. 9, where $R_s$ is called $R_{BCS}$). Superimposed are the results from eqs. (1) and (3), obtained with the parameters as in Table 3 (coloured lines) for 120°C baked cavities.

The relatively small value of $B_c^*$ of the N-doped cavity (Table 1), in comparison to that of the BCP cavity (Table 3) needs an explanation. In the framework of the model presented this difference can be attributed to a smaller depth $d^*$ of the accountable N-doping owing to the relatively large and sharply distributed volume fraction $x_1=0.667$ of the nitrogen component.

From Table 3 one concludes that the data are fitted with essentially pretty similar parameters as those of Figure 1, except for the saturation field $B_c^*$.

Table 3: Parameters used for fitting the data of Figure 4

| | | |
|---|---|---|
| $x_1$ | 0.667 | A priori parameters |
| $B^*$ [mT] | 16 | |
| $l$ [nm] | 22*) | Derived parameters |
| $\lambda$ [nm] | 64*) | |
| $d^*$ [nm] | 147±11 | |
| $s_{Nb}$ [1/Ωm] | $5.1 \cdot 10^{7*)}$ | |
| RRR | 7.6±0.4 | Free fit parameters |
| $B_c^*$ [mT] | 156±31 | |
| *) The error is smaller than the last digit indicated | | |

## CONCLUSION

In this paper, the model as proposed in ref. 6 was challenged with new data. The model is extended by taking into account the complete formula of Bruggeman/Landauer describing the electrical conductivity of a binary uniform metallic mixture, represented by a weak superconductor. The new data can be explained by an imaginary electrical conductivity of the "niobium" component of the weak superconductor. By virtue of eq. (1), the model shows quite naturally, why the field dependence of the surface resistance is often attributed to $R_{BCS}$. The consistency with the superconducting proximity effect of the obtained fitting results is confirmed. It is conjectured that the doping agent is not the most relevant parameter for the increase of Q with $E_{acc}$, but rather percolation effects. The dependence of Q on $E_{acc}$ is observed not only for surfaces in cavities doped with nitrogen but also for undoped chemically polished and baked ones. It is also worth mentioning that doping with other elements than nitrogen (e.g. Ar) may as well lead to an increase of Q with $E_{acc}$ [7], though not across all laboratories [23]. In the latter case, though, it is likely that little to no argon diffused into the niobium during the doping process. This observation may point towards the doping agent being less or even not important for achieving a gain of Q, but rather geometrical features, such as percolation, inside a disordered and sufficiently low $T_{cN}$- composite. A similar conjecture was already proposed elsewhere for non-doped cavities after chemical or electrical polishing [24].

## REFERENCES

[1] A. Gurevich, Multiscale mechanisms of SRF breakdown, Physica **C 441** (2006) 38.
[2] B. Xiao and C. E. Reece, A new first-principles calculation of field-dependent RF surface impedance of BCS superconductor and application to SRF cavities, arXiv:1404.2523.
[3] A. Gurevich, Reduction of dissipative nonlinear conductivity of superconductors by static and microwave magnetic fields, Phys. Rev. Lett. **113** (2014) 087001.
[4] W. Weingarten, R. Eichhorn, N. Stillin, N-doped niobium accelerating cavities: Analyzing model applicability, Proceedings of LINAC2016, East Lansing, MI, USA, p. 1014.
[5] J. T. Maniscalco, M. Liepe, D. Gonnella, The importance of the electron mean free path for superconducting RF cavities, Journ. App. Phys. **21** (2017) 043910.
[6] R. Eichhorn, D. Gonnella, G. Hoffstaetter, M. Liepe, and W. Weingarten, On superconducting niobium accelerating cavities fired under $N_2$-gas exposure, arXiv:1407.3220.

[7] A. Romanenko and A. Grassellino, Dependence of the microwave surface resistance of superconducting niobium on the magnitude of the RF field, Appl. Phys. Lett. **102** (2013) 252603.

[8] A. Grassellino, A. Romanenko, D. A. Sergatskov, O. Melnychuk, Y. Trenikhina, A. Crawford, A. Rowe, M. Wong, T. Khabiboulline, F. Barkov, Nitrogen and argon doping of niobium for superconducting radio frequency cavities: a pathway to highly efficient accelerating structures, Supercond. Sci. Technol. **26** (2013) 102001.

[9] M. Martinello, S. Aderhold, S. K. Chandrasekaran, M. Checchin, A. Grassellino, O. Melnychuk, S. Posen, A. Romanenko, D.A. Sergatskov, Advancement in the understanding of the field and frequency dependent microwave surface resistance of niobium, arXiv:1707.07582v1.

[10] G. Linker, Superconducting properties and structure of ion bombarded Nb layers, Radiation Effects **47** (1980) 225.

[11] G. Linker, Superconductivity in d- and f-Band Metals, W. Buckel, W. Weber (eds.), Kernforschungszentrum Karlsruhe, Germany, 1982, p. 367.

[12] J. Spitz, J. Chevallier, A. Aubert, Propriétés et Structure des couches minces de nitrure de niobium élaborées par pulvérisation cathodique réactive, Journ. Less-Common Metals **35** (1974)181.

[13] Y. Trenikhina, A. Grassellino, O. Melnychuk, A. Romanenko, Characterization of niobium doping recipes for the Nb SRF cavities, Proceedings SRF2015, Whistler, BC, Canada, p. 223.

[14] R. Landauer, The electrical resistance of binary metallic mixtures, Journ. Appl. Phys. **23** (1952) 779.

[15] B. I. Halperin, D. J. Bergman, Heterogeneity and Disorder: Contributions of Rolf Landauer, arXiv:0910.0993v1.

[16] D. A. G. Bruggeman, Ann. Phys. (Leipzig) **24** (1935) 636.

[17] E. Schröder, Über supraleitende Verbindungen des Niob, Superconductive Compounds of Niobium, Z. Naturforschg. **12a** (1957) 247.

[18] P. G. de Gennes and J. P. Hurault, Proximity effects under magnetic fields II - Interpretation of "breakdown", Phys. Lett. **17** (1965) 181.

[19] E. Krätzig and W. Schreiber, Ultrasonic Investigation of Proximity Effects in Superconductivity under the Influence of Magnetic Fields, Phys. kondens. Materie **16** (1973) 95.

[20] A. T. Fiory, B. Serin, Thermomagnetic properties of the mixed state, Physica **55** (1971) 73.

[21] B.B. Goodman and G. Kuhn, Influence des défauts étendus sur les propriétés supraconductrices du niobium, J. Phys. Paris **29** (1968) 240.

[22] A. F. Mayadas, R. B. Laibowitz, and J. J. Cuomo, Electrical characteristics of rf-sputtered single-crystal niobium Films, J. Appl. Phys. **43** (1972) 1287.

[23] P. N. Koufalis, F. Furuta, M. Ge, D. Gonnella, J. J. Kaufman, M. Liepe, J. T. Maniscalco, Impurity doping of superconducting RF cavities, Proceedings of IPAC2016, Busan, Korea, p. 3195.

[24] W. Weingarten, Field-dependent surface resistance for superconducting niobium accelerating cavities, Phys. Rev. STAB **14** (2011) 101002.